\documentclass{iau}
\usepackage{graphicx}

\title[IAUS291.~~FAST low frequency pulsar survey] 
{FAST low frequency pulsar survey} 

\author[Y. Yue, D. Li \& R. Nan]  
{Youling Yue, Di Li \& Rendong Nan 
}

\affiliation{
National Astronomical Observatories, Chinese Academy of Sciences\\
20A Datun Road, Chaoyang District, Beijing 100012, China
\\email: {\tt ylyue@nao.cas.cn} \\[\affilskip]
}

\pubyear{2012}
\volume{291}  
\jname{\mbox{Neutron Stars and Pulsars: Challenges and Opportunities after 80 years}}
\editors{J. van Leeuwen, ed.}
\begin{document}

\maketitle

\begin{abstract}
The Five-hundred-meter Aperture Spherical radio Telescope (FAST) is under construction and will be commissioned in September 2016.
A low frequency 7-beam receiver working around 400 MHz is proposed for FAST early science.
It will be optimized for a whole FAST sky drift-scan pulsar survey.
Simulations show that about 1500 new normal pulsars will be discovered, as while as about 200 millisecond pulsars.
\end{abstract}


\firstsection 
\section{Introduction}

The Five-hundred-meter Aperture Spherical radio Telescope (FAST) is now under construction.
The first light is expected to be in September 2016
\cite[(Nan et al.\ 2011)]{Nan2011}.
At the early science phase, a multibeam receiver working at low frequency ($<1$ GHz),
which has less stringent pointing accuracy requirement, is favored.
Such a receiver will even work during the testing phase.

Besides the current 8 sets of receivers, a 7-beam receiver is planned for FAST to do an early pulsar survey.
It will work at around 400 MHz with a bandwidth about 150 MHz,
which is optimized for whole FAST sky drift-scan pulsar survey.
Because the pulsar flux drops down at higher frequency and receiver has smaller beam size at higher frequency,
the L-band 19-beam receiver is not sufficient for a whole FAST sky pulsar survey.
The 7-beam receiver will be the best option to do this
before the Phase Array Feed  receiver is available for FAST.
This receiver can also be used for a high redshift neutral hydrogen survey and other sciences.

To minimize the risk and to make sure the 7-beam receiver is ready in
early 2016, we will use a mature dipole design or horn design like current multibeam receivers such as the Parkes 13-beam and the Arecibo 7-beam receivers.

Outlined below are some tentative technical specifications of the receiver:
\begin{itemize}
\item
Frequency $f\sim400$ MHz, bandwidth $\sim1/3f$, which is feasible by available technology
\item
System temprature without sky $T_{\rm sys}\sim30$ K or less, cooled with lightweight Stirling refrigerator
\item
Use horn or dipole design
\item
Light weight, easy to manufacture and inexpensive
\item
Use 19-beam receiver backends
\end{itemize}

The 7-beam receiver will be optimized for pulsar and transient survey at early science phase or even earlier.
Three probable surveys are listed below.
\begin{itemize}
\item Low frequency drift-scan pulsar survey:
Detect $\sim2300$ normal pulsar ($\sim1500$ new).
Detect $\sim300$ millisecond pulsars (MSPs) ($\sim200$ new), which is good for gravitational wave detection.
Data of one whole FAST sky scan $\sim2.4$ petabytes.
Simulation details are described in Section 2.
\item M31/M33 pulsar survey:
Probably first detection of extragalactic pulsar beyond Magellanic Clouds.
\item Radio transient survey:
Piggyback survey of low frequency drift-scan pulsar survey. Use same data set.
Scan the whole FAST sky a few times during early science phase.
\end{itemize}

\section{Pulsar survey simulation}
To find the best frequency for the 7-beam receiver,we have done survey simulation
using PSRPOP (http://psrpop.sourceforge.net) \cite[(Lorimer et al.\ 2006)]{Lorimer2006}.
The pulsar population generation is similar to Smits et al. (2009).

In drift-scan mode, the integration time is decided by the beam width which is inversely proportional to observing frequency. It is about 40 seconds at 400 MHz. The survey speed also depends on frequency. At 400 MHz, the whole FAST sky (2.3$\pi$) will be covered in 2 months.
Two working case are considered.
\begin{itemize}
\item Spherical surface:
The illuminated aperture decreases as frequency increases, $D_{\rm ill}\sim 200\times(f/400{\rm ~MHz})^{1/4}$ m \cite[(Condon 1969)]{Condon1969}. This will be the case at the very early phase before early science phase, when the reflector has just been laid.
\item 300 meter diameter parabola surface:
The illuminated aperture is a 300 meter diameter parabola.
This will be the case at the early science phase.
\end{itemize}

Since normal pulsars and MSPs constitute different populations, they
are discussed separately. First the normal pulsar population is discussed in detail. Results are shown in Fig. 1 and 2.

\begin{figure}[!htb]
\begin{center}
\includegraphics[width=10cm,angle=0]{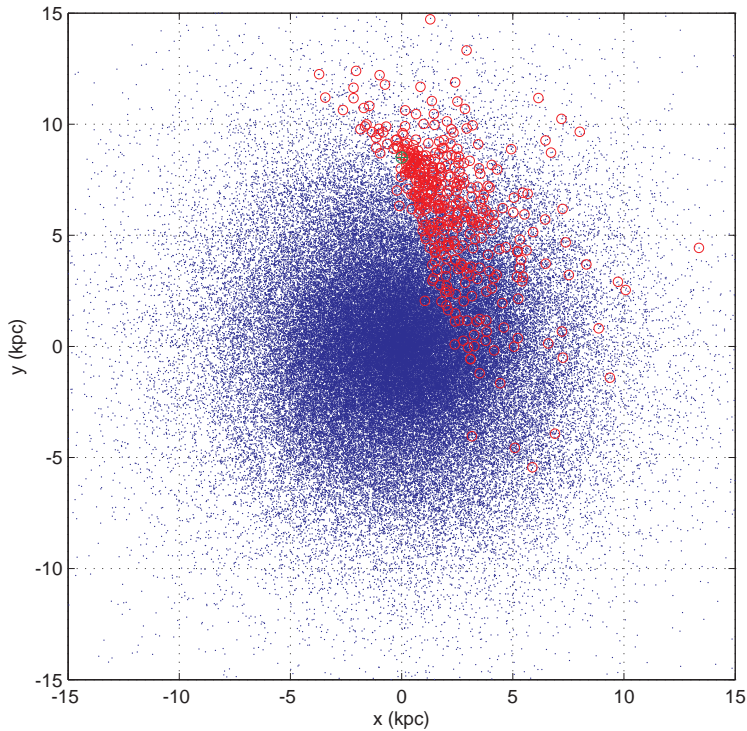}
\caption{
Output of one simulation run.
Dots are the $\sim100$ thousand normal pulsar generated.
Circles are the $\sim2300$ pulsars detected by FAST at 400 MHz band, $\sim1500$ would be new.  Bandwidth = 1/3  central frequency is assumed. Reflector surface is 300 meter diameter parabola.
}
\label{fig1}
\end{center}
\end{figure}

\begin{figure}[!htb]
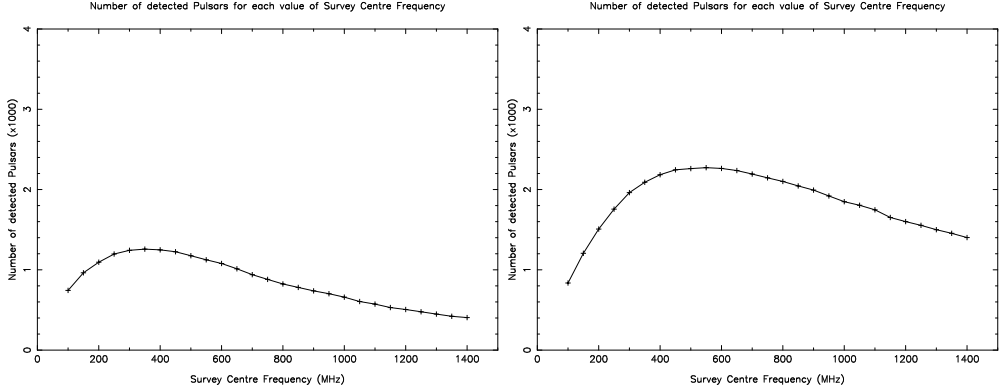

\begin{center}
\includegraphics[width=5.1cm,angle=-90]{300m-sph}
\includegraphics[width=5.1cm,angle=-90]{300m-para}

\caption{
Number of normal pulsars detected at different frequencies. Bandwidth is
1/3rd of center frequency and 40\,s integration is assumed.
{\it Left:} for spherical reflector surface. The illuminated aperture
decreases with frequency , $D_{\rm ill}\sim 200$ m$\times(f/400{\rm ~MHz})^{1/4}$ \cite[(Condon, 1969)]{Condon1969};
A maximum of 1200 pulsars will be detected.
{\it Right:} for reflector with a 300\,m-diameter parabolic illuminated surface.
A maximum of 2300 pulsars will be detected.
Since there are already $\sim800$ known pulsars in the FAST sky,
a total of 1500 new normal pulsars are expected to be discovered.
The number of detected pulsars peaks around 500 MHz,
but does not vary much when the center frequency is in the range 400 to 700 MHz.
Considering number of pulsars detected and survey speed, the lower frequency end 400 MHz is favored.
}
\label{fig2}
\end{center}
\end{figure}

\section{Discussion}

MSPs are different to normal pulsars, having e.g. a different spectral
index, and spatial distribution. A spectral mean $-1.6$ with deviation 0.35 is used by Smits et al. (2009).
We find that using these values will underestimate MSPs detected in the Parkes 70 cm survey.
We then treat spectral index and spectral index deviation as free parameter, and find in the parameter space where the simulation agrees with both Parkes multibeam survey and 70 cm survey. The result favor steeper spectral index or larger spectral index deviation.
Details will be presented in a later paper.

A pulsar survey is sensitive to RFI, which should be carefully considered.
The overall RFI situation at FAST site is good around 400 MHz.
Only a few narrow-band RFI instances exist. The final frequency and
bandwidth of the receiver will be decided after new measurements of the RFI.

\section*{Acknowledgments}
The 7-beam receiver was initially proposed by Jim Condon.
We appreciate the valuable comments and suggestions from the pulsar community and FAST group,
and help from Fredrick Jenet, Duncan Lorimer and Roy Smits.
This work is supported by
the National Natural Science Foundation of China (11103045)
and by
China Ministry of Science and Technology under State Key
Development Program for Basic Research (2012CB821800).

\end{document}